\documentclass[aps,prl,twocolumn,superscriptaddress]{revtex4-1}

\usepackage{graphicx}

\begin{document}


\title{Turbulent dynamics of epithelial cell cultures}



\author{C. Blanch-Mercader}
\email{carles.blanch-mercader@curie.fr}
\affiliation{Laboratoire PhysicoChimie Curie, Institut Curie, PSL Research University - Sorbonne Universit\'es, UPMC?CNRS - Equipe labellis\'ee Ligue Contre le CancerÊ; 75005, Paris, France.}

\author{V. Yashunsky}
\email{Victor.Yashunsky@curie.fr}
\affiliation{Laboratoire PhysicoChimie Curie, Institut Curie, PSL Research University - Sorbonne Universit\'es, UPMC?CNRS - Equipe labellis\'ee Ligue Contre le CancerÊ; 75005, Paris, France.}

\author{S. Garcia}
\affiliation{Laboratoire PhysicoChimie Curie, Institut Curie, PSL Research University - Sorbonne Universit\'es, UPMC?CNRS - Equipe labellis\'ee Ligue Contre le CancerÊ; 75005, Paris, France.}

\author{G. Duclos}
\affiliation{Laboratoire PhysicoChimie Curie, Institut Curie, PSL Research University - Sorbonne Universit\'es, UPMC?CNRS - Equipe labellis\'ee Ligue Contre le CancerÊ; 75005, Paris, France.}

\author{L. Giomi}
\affiliation{Instituut-Lorentz, Universiteit Leiden, P.O. Box 9506, 2300 RA Leiden, Netherlands}



\author{P. Silberzan}
\affiliation{Laboratoire PhysicoChimie Curie, Institut Curie, PSL Research University - Sorbonne Universit\'es, UPMC?CNRS - Equipe labellis\'ee Ligue Contre le CancerÊ; 75005, Paris, France.}



\date{\today}

\begin{abstract}
In vitro epithelial monolayers exhibit a rich repertoire of collective dynamical behaviours according to microenvironment mechanical features. We investigate the statistical and dynamical properties of human bronchial epithelial cell (HBEC) flows in bulk, by analysing their large length-scale and long time-scale spatiotemporal dynamics. Activity-driven spontaneous collective flows consist of an ensemble of vortices randomly positioned in space. By analysing a large population of vortices, we show that their area follows an exponential law with a constant mean value and their rotational frequency is size-independent, both being characteristic features of the chaotic dynamics of active nematic suspensions. Indeed, we find that HBECs self-organise in nematic domains of several cell lengths. Nematic defects are found at the interface between domains with a total number that remains constant
by the dynamical balance of nucleation and annihilation events.
Near $\pm 1/2$ defects, the mean velocity field is in faithful agreement with the dynamics of extensile active nematics. 
\end{abstract}


\maketitle


In vitro epithelial monolayers have become a model system to investigate fundamental processes in living organisms, such as morphogenesis \cite{lecuit2007orchestrating,bryant2008cells,friedl2009collective} or regeneration \cite{sonnemann2011wound,cochet2014border,brugues2014forces}. The collective dynamics of cellular systems has gathered increasing attention over the last few decades, inspiring experimental and theoretical studies in an attempt to unravel their underlying physical principles \cite{hakim2017collective}. Single cells are able to convert stored chemical energy into mechanical forces \cite{du2005force} and modify accordingly their local microenvironment \cite{bendix2008quantitative}. At the multicellular level, these self-generated forces can be transmitted over tens of cell lengths \cite{trepat2009physical}, resulting in cellular motion throughout the entire system which self-organizes into a vast range of flowing patterns \cite{poujade2007collective,angelini2011glass,vedula2012emerging,serra2012mechanical,doxzen2013guidance,deforet2014emergence,vedula2014epithelial}. Remarkably, mesoscopic cellular behaviour ($\sim$ few hundreds of microns) is known to be influenced by physical confinement. For instance, constrained ensembles of elongated cells develop spontaneously antiparallel shear flows \cite{duclos2017shear}.
MDCK epithelial cells exhibit qualitatively different modes of migration upon variations of the confinement's geometry \cite{poujade2007collective,vedula2014epithelial}. However much less work has been put forward to characterise the emergent macroscopic behaviour of weakly-adherent spindle-shaped cells in vitro ($\sim$ few thousands microns), despite its importance in biophysics. 

The large-scale behaviour of highly active epithelial monolayers often features signatures of erratic motion \cite{deforet2014emergence,vedula2014epithelial}, however whether these can be attributed solely to the intrinsic cellular stochasticity, or  they are a manifestation of some form of collective modes in a turbulent regime still remains an open question. Remarkably other biological systems in similar physical conditions such as bacterial colonies or suspensions of cytoskeleton's filaments and molecular motors \cite{schaller2010polar,dunkel2013fluid,keber2014topology}, seem to exhibit emergent spatio-temporal chaotic patterns for which formation and annihilation of mesoscopic swirls and jets occurs. Despite differing in length, time or energy scales, these systems are often faithfully described by continuum models \cite{marchetti2013hydrodynamics}. 
From a theoretical point of view, it is customary to build simple approaches by disregarding the specific microscopic details of the units, to shed light on the generic mechanisms  connecting the material properties of the active units to their macroscopic behaviour. Turbulence in active systems evidenced in Toner-Tu models \cite{wensink2012meso,dunkel2013fluid}, generalized Swift-Hohenberg models \cite{oza2016generalized} and active nematic gels \cite{thampi2014vorticity,giomi2015geometry,ramaswamy2016activity,genkin2017topological} are often mediated by the dynamics of topological nematic defects. Alternative routes to chaotic dynamics in polar systems have been identified \cite{marcq2014,blanch2017hydrodynamic}. Although cell cultures share common symmetries with the previously mentioned biological examples, no chaotic dynamics was reported, possibly because of the high cell-substrate friction forces. 

In the present Letter, we report the existence of a turbulent regime in human bronchial epithelial cells (HBEC) over three decades in kinetic energy,  and we show that our findings are in agreement with an hydrodynamical model of an active nematic fluid.

We study experimentally the nematodynamics of immortalized HBEC monolayers and its evolution over time. Conversely to other epithelial cell lines, HBECs are weakly cohesive at initial stages, meaning that when cells fully cover the substrate, they are still highly motile and exhibit long range collective movements. During the next few tens of hours, the system gradually slows down approaching asymptotically a jammed state, in which cells hardly move beyond their own sizes  \cite{garcia2015physics}. Over the course of the experiments ($\sim 60$ hours), the speed of the population drops more than one decade, while maturation of cell-cell and cell-substrate adhesions occurs \cite{garcia2015physics}.

 \begin{figure}[t]
 \includegraphics[width=8.7cm]{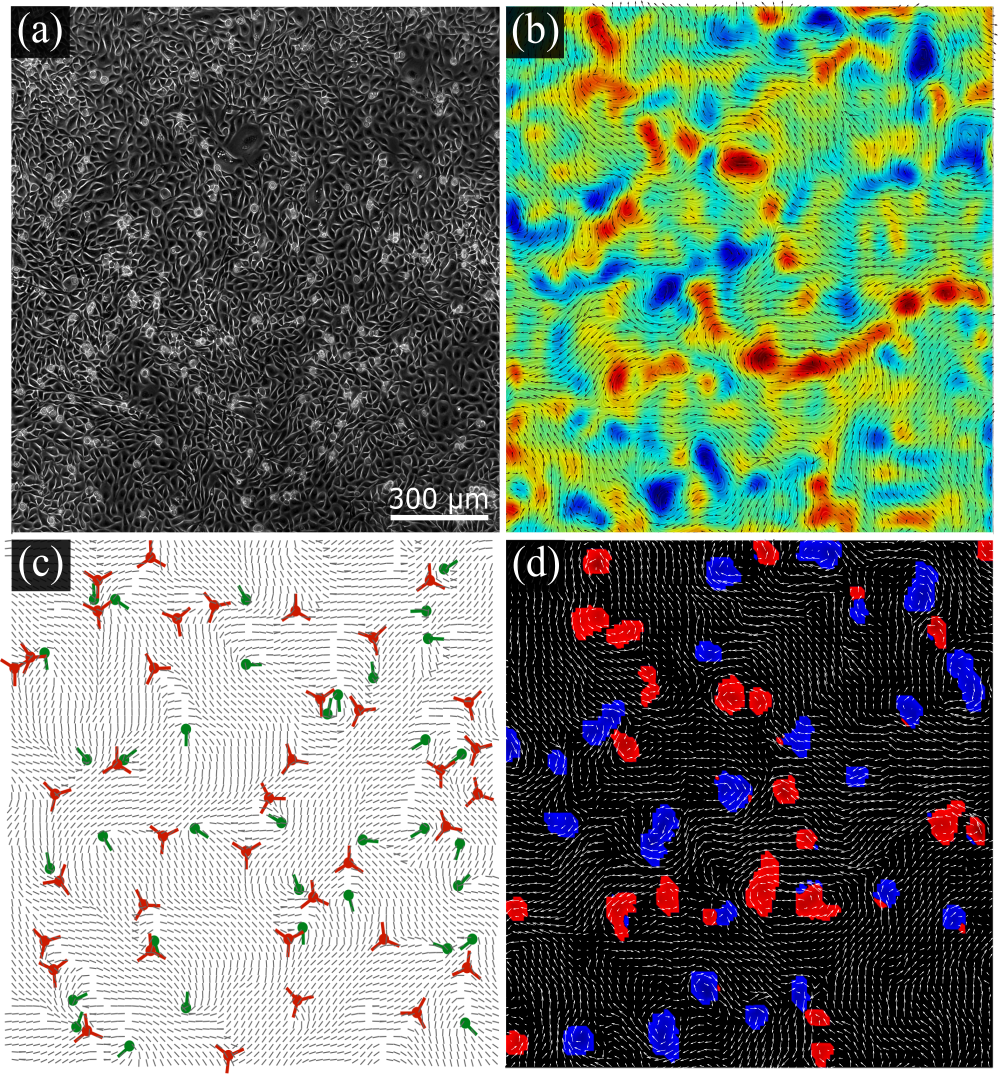}%
 \caption{Nematodynamics of HBECs. (a) Phase-contrast image. (b) Normalized vorticity map with the PIV flow field showing collective behaviour. (c) Cell Orientation map with green (red) dots marking the position of $+1/2$ ($-1/2$) defects. (d) Map of the Okubo-Weiss field thresholded to negative values (i.e. $Q<0$ in the text) where blue (red) domains indicate CW (CCW) rotating vortices. Scale bar is $300$~$\mu$m. \label{fig::1}}
 \end{figure}
 
At intermediate time regimes ($\sim 20$ hours) HBECs form a cohesive monolayer, showing large-scale collective movements Fig.~(\ref{fig::1},a). 
Mapping the velocity field $(v_{x},v_{y})$ \cite{petitjean2010velocity}, we observe large and transient vortices, which show no apparent sign of organization Fig.~(\ref{fig::1},b).
To quantify the global properties of cell flows, we define the in-plane enstrophy $\Omega=\langle \omega^2/2 \rangle$ as the square of the $2d$ vorticity field $\omega=\partial_{x}v_{y}-\partial_{y}v_{x}$ and the in-plane kinetic energy per unit mass $\mathcal{E}=\langle (v_{x}^2+v_{y}^2)/2 \rangle $, where the symbol $\langle \cdot \rangle$ denotes spatial averages. Both are standard quantities in $2d$ turbulence studies, \cite{dunkel2013fluid}. Their ratio $\mathcal{E}/\Omega$ has units of length square and it can be related to the mean vortex radius \cite{dunkel2013fluid}, which in our case is of the order of $\sim 50$~$\mu$m. The vortex area is quantified using the Okubo-Weiss criterion \cite{Okubo1970,WEISS1991}, which identifies the elliptic regions around vortices for which the field $Q=(\partial_{y}v_{x}\partial_{x}v_{y}-\partial_{x}v_{x}\partial_{y}v_{y})<0$ as shown in Fig.~(\ref{fig::1},d). Notice that the Okubo-Weiss field can be recasted as $Q=-\det \mathbf{\nabla}\mathbf{v}$, meaning that the previous condition enforces that the eigenvalues of the velocity gradient tensor are purely imaginary conjugates for incompressible flows  \cite{Okubo1970,WEISS1991}.

When crowded, motile HBECs deform and acquire elongated shapes. The anisotropy of their shapes permit to assign a local nematic orientation, which is described by the unit vector $(\hat{n}_x,\hat{n}_y)$. Remarkably, cells tend to self-organize into transient mesoscopic domains with a well-defined nematic order, separated by pairs of $\pm 1/2$ defects Fig.~(\ref{fig::1},c). 
The orientation of $+1/2$ ($-1/2$) defects is marked by a green (red) symbols, following Ref.~\cite{vromans2016orientational}. Recent works have shown 
that the defect position correlates with MDCK cell extrusion and progenitor cell aggregation~\cite{saw2017topological,kawaguchi2017topological}. Notice that the measured nematic order is dependent upon the cellular shape alone, which in turn might be controlled by the force generation machinery in a non-trivial manner as shown for vertex models \cite{czajkowski2017hydrodynamics}. 

\begin{figure}[t]
	\includegraphics[width=8.7cm]{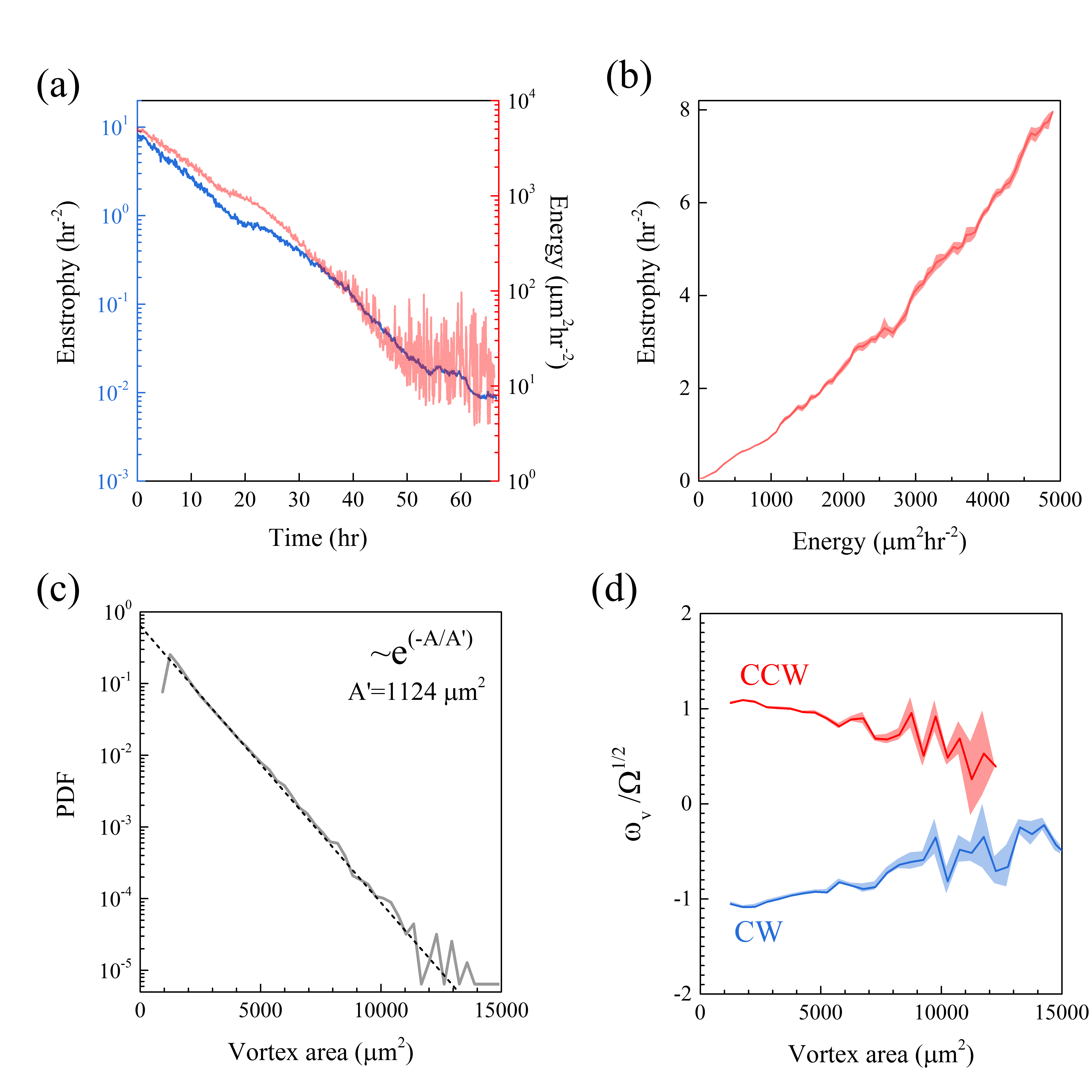}%
	\caption{Statistical analysis of the dynamics of mesoscopic cell flows. (a) Log-linear representation of the time evolution of enstorphy $\Omega$ and kinetic energy per unit mass $\mathcal{E}$. (b) Linear scaling between $\Omega-\mathcal{E}$ over the entire experiment. (c) Log-linear presentation of the total vortex area distribution. The dotted line is an exponential fit. (d) Total mean vortex angular velocity normalised by the square root of the enstrophy $\Omega$ as a function of the vortex area.  \label{fig::2}}
\end{figure}

Over the course of the experiment, cell density increases significantly by proliferation, influencing drastically cell behaviour \cite{garcia2015physics}. We observe that the global properties of the cell flows vary in time in an exponential manner over a range that spans three decades in enstrophy Fig.~(\ref{fig::2},a). The inverse of the squared root of the enstrophy defines a reference time scale for the instantaneous dynamics of the system. During the first $40$~hours, $\Omega^{-1/2}$ varies continuously between $0.1$ to $1$ hours. However due to density variations, the dynamics itself 
slows down with a typical time scale of $\partial_{t}\Omega/\Omega\sim 10$~hour. The separation of these two time scales suggest that the material properties of the system are changing in an adiabatic manner \cite{garcia2015physics}. 
As the enstrophy is a monotonic function of time, from now on we map the temporal coordinate into $\Omega^{-1/2}$ to average over several experiments.  
 
Based on our findings, we hypothesize that HBEC monolayers behave as an active nematic fluid, \cite{giomi2015geometry}.
Active matter describes generically assemblies of units with an internal source of chemical energy in non-equilibrium conditions, such as bacterial suspensions or cytoskeleton model systems \cite{Kruse2005,Julicher2007}. 
It has been shown theoretically that above a certain critical level of activity, 
active nematic suspensions develops $2d$ turbulent flows characterised by the nucleation and annihilation of defect pairs. The interplay between the nematic defects and activity gives rise to the spontaneous formation of transient swirls.
A reduced set of parameters suffice to draw several scaling laws of their flow properties, which are the Frank constant of the nematic suspension $K$, the shear viscosity $\eta$ and the scale of the anisotropic active stresses $\alpha$. The former is negative (resp. positive) for extensile (resp. contractile) particles. For further details on the model, we refer the reader to Ref.~\cite{giomi2015geometry}.

The main features defining turbulence in active nematics are the exponential distribution of vortex area and the uniform distribution of the mean vorticity per vortex \cite{giomi2015geometry}. A set of scaling laws ensue from these statistical properties such as the average vortex radius size $\ell_{a}\sim \sqrt{K/|\alpha|}$, which arises from the competition between the energetic cost to form spatial distortions of the nematic field and the intrinsic activity of the system. Our measurements Fig.~(\ref{fig::2},b) indicate that the energy-enstrophy ratio remains approximately constant over the entire evolution, suggesting that the time-variation of $K$ and $\alpha$ are very limited. A detailed look at Fig. (\ref{fig::2},b) shows that the curve $\Omega(\mathcal{E})$ is slightly convex, meaning that the average vortex radius increases from  $44\pm 2$ $\mu$m to $55\pm 2$ $\mu$m with the decrease of the kinetic energy per unit mass. This observation is compatible with the prediction that higher activity results into smaller vortices \cite{thampi2014vorticity,giomi2015geometry}. 
At a given time, the individual vortex area detected with Okubo-Weiss criterion exhibits an exponential distribution for a broad range of areas. We verified that the form of the distribution function remains constant over the entire experiment. 
Hence a temporal average over a population of about $10000$ independent vortices confirms and extends our initial observation that the vortex area is exponentially distributed with a mean vortex area of $1120$ $\mu$m$^2$ Fig.~(\ref{fig::2},c), which is of the order of the scale provided by squared energy-enstrophy ratio. On the other side, cell flows scale as the vorticity, which in turn obeys the scaling law $\omega\sim\alpha/\eta$ for an active nematic. Assuming that the activity remains constant, Fig.~(\ref{fig::2},a) suggests that the shear viscosity varies more than a decade. As it has been shown, cell density increases steadily over time and the system eventually approaches a jammed phase, \cite{garcia2015physics,sadati2013collective}. Theoretical studies of ensembles of self-propelled particles has shown that the mean squared displacement of single particles is strongly controlled by system density \cite{henkes2011active,ni2013pushing,berthier2014nonequilibrium}. However in our system other density-independent mechanisms might be also relevant as cell-cell connectivity has been shown to influence HBEC flows \cite{garcia2015physics}. At all time-points, we measure the mean vorticity per vortex $\omega_v$ and we find that it scales with the square of the enstrophy $\sim \Omega^{1/2}$ Fig.~(\ref{fig::2},d). 
We observe that the total distribution of the normalized mean vorticity $\omega_v/\Omega^{1/2}$ is weakly dependent of the vortex area and the rotation of vortices is evenly distributed between CW and CCW, at the accuracy of the experiments Fig.~(\ref{fig::2},d). In conclusion we show that 
for a statistical sample of more than $10000$ independent vortices, the distribution of sizes obeys an exponential law and that the vortex rotational speed is approximately constant with a slight tendency to decrease with the vortex area, both features being strong signatures of turbulence in active nematic. 

 \begin{figure}
 \includegraphics[width=8.7cm]{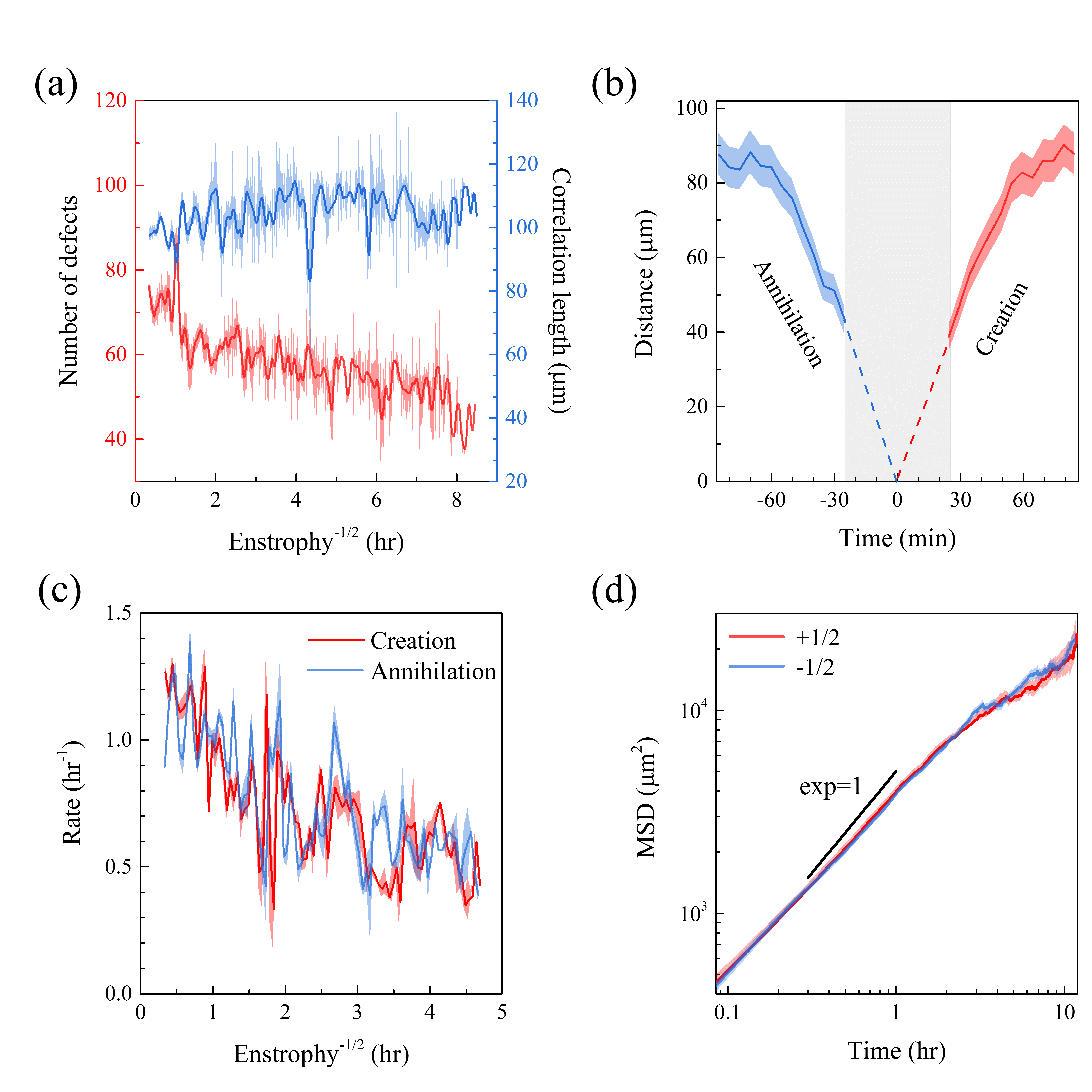}%
 \caption{Statistical analysis of the cell nematic order parameter. (a) Number of defects and nematic Correlation length as a function of $\Omega^{-1/2}$. We employ the enstrophy to label the time coordinate. (b) Small-time dynamics of the minimal distance between two opposed nematic defects close to the nucleation (annihilation) in red (blue). (c) Time evolution of the creation/annihilation rates as a function of $\Omega^{-1/2}$. (d) Diffusive behaviour on the long-time dynamics of nematic defects, where MSD curves are fitted by $\sim D t^{n}$.  \label{fig::3}}
 \end{figure}
 
To further test our hypothesis, we analyse the statistical properties of cell orientation. At a given time, HBECs tend to form transient finite-size domains with high orientational order Fig.~(\ref{fig::1},c). The typical domain size is given by the nematic correlation length $\ell_c$, or the decaying length of the director correlation function ${\cal{C}}_n=\langle\mathbf{\hat{n}}(x'+x,y'+y)\cdot\mathbf{\hat{n}}(x',y')\rangle_{(x',y')}$, where the spatial average is over the entire field of view and $\mathbf{\hat{n}}$ denotes cell orientation. At the interface between these domains, $\pm 1/2$ defects are found, which are a hallmark of nematic liquid crystals \cite{de1993physics}. These singular structures were previously found in epithelial, fibroblast, myoblast and progenitor cells \cite{kemkemer2000elastic,duclos2017topological,saw2017topological,kawaguchi2017topological} as well as in other biological systems \cite{schaller2013topological,decamp2015orientational,genkin2017topological}. Strikingly, in our system the total number defects decrease by less than $15\%$ upon a $100$-fold variation of the kinetic energy per unit mass Fig.~(\ref{fig::3},a). We found that $\ell_c\sim 100$ $\mu$m is comparable to the vortex size, a regime in which the total number of defects is activity-independent for active nematics \cite{giomi2015geometry}. 
However unlike in other cell monolayers \cite{duclos2017topological}, the number of defects is set by the dynamic balance between the creation and annihilation events of $\pm 1/2$ defects pairs. The typical time scale of these processes is about $\sim 1$~hour and it increases as the total enstrophy rises Fig.~(\ref{fig::3},c).
Analyzing the trajectories of individual defects at long-time scales (more than $100$~min), we observe that the $2d$ mean squared displacement follows a power law with an exponent close to $1$, yielding an effective diffusion coefficient of $\sim (34\pm 2)*$~$10^2 \mu$m$^2$/hr with no apparent quantitative differences between both type of defects Fig.~(\ref{fig::3},d). However at shorter time scales their dynamics exhibit distinct and remarkable features. Fig.~(\ref{fig::3},b) shows the trajectories of the minimal distance between two opposite-charge defects $\Delta$ toward the annihilation (blue) or away from the nucleation (red) event, where the time origin has been set at annihilation or creation instant. Their trajectories
have opposite but symmetrical profiles (Fig.~\ref{fig::3},b), saturating over distances longer than the nematic correlation length $\ell_c$. The resolution of our analysis does not allow resolving the dynamics close to the core of the defects. Despite the collective nature of turbulence in active nematics, it is theoretically demonstrated that the distance $\Delta$ between defects with opposed sign obeys a local overdamped equation of motion of the form $\partial_t \Delta=\pm v_0-{\mathcal{\kappa}}/\Delta$ up to higher order corrections in $\Delta$, where the positive (negative) sign applies to the nucleation (annihilation) case \cite{pismen2013dynamics,giomi2014defect}. The coefficient $v_0$ stands for the self-propeled velocity of $+1/2$ defects and $\mathcal{\kappa}$ is related to the nematic properties of the active units. By fitting them on our trajectories from Fig.~(\ref{fig::3},b) we obtain that $v_0=1.3\pm 0.2$~$\mu$m/min and $\mathcal{\kappa}<6$~$\mu$m$^2$/hr, suggesting that the elastic attractive interaction between defects is subdominant against the activity of $+1/2$ defects for HBEC monolayers.

 \begin{figure}
 \includegraphics[width=8.7cm]{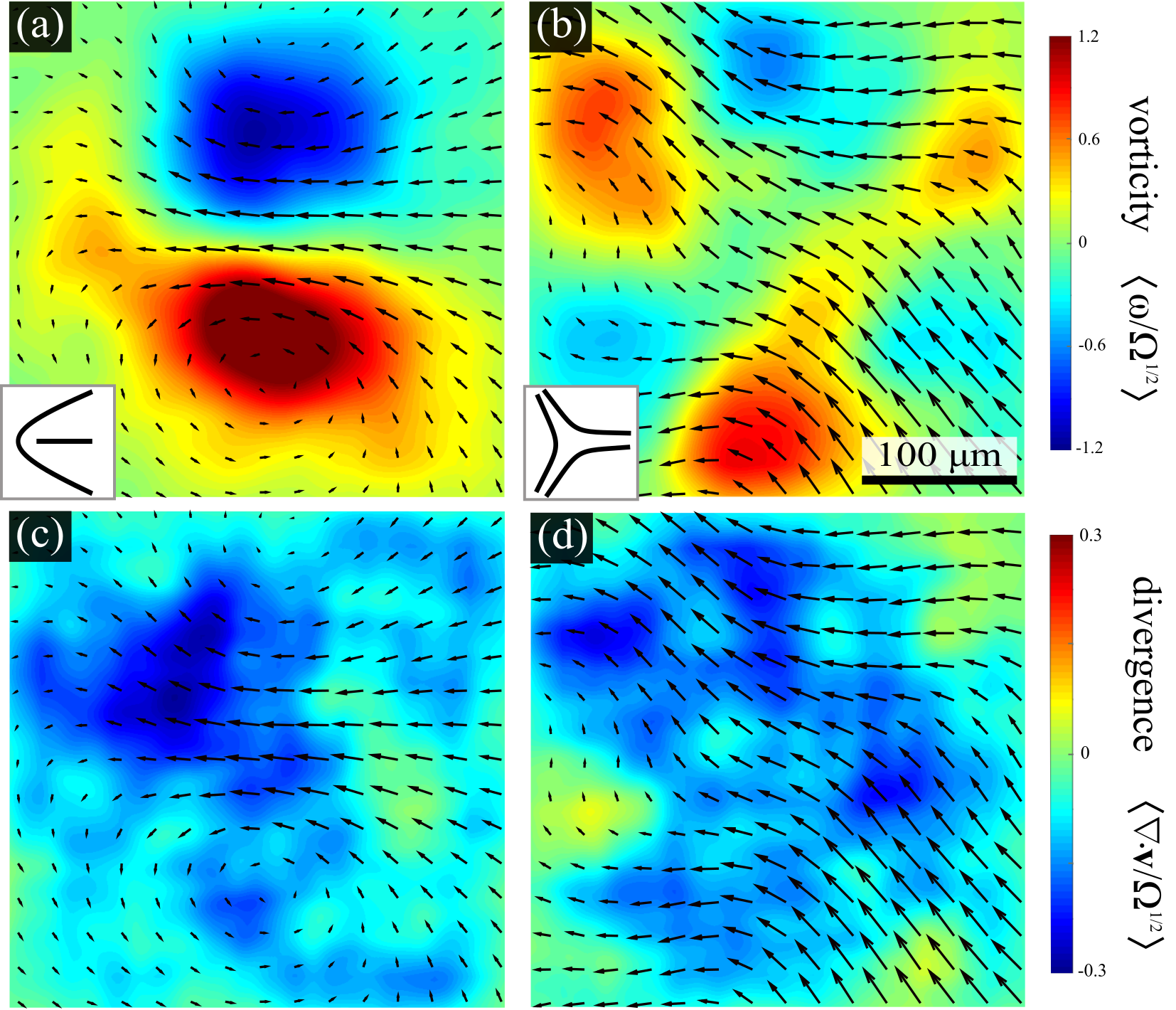}%
 \caption{Ensemble average of the flows near nematic defects. Left (Right) column shows the normalised by their maximal value vorticity and divergence of the velocity field for a $+1/2$ ($-1/2$) defect. The overlaid white arrow represent the direction of the velocity field for each case. The upper panel insets are sketches of the nematic orientation around the defect. \label{fig::4}}
 \end{figure}
 
To further explore the dynamics of nematic defects, we analyse the local properties of cell flows around the singularities of the nematic order field. Hence we compute the local mean velocity field for a population of $\sim 10000$ nematic defects, Fig.~(\ref{fig::4}). The scale of the velocity is normalised to the instantaneous value of $\Omega^{1/2}$. The vorticity map of $+1/2$ defects presents a symmetrical organisation of a pair of counter-rotating vortices Fig.~(\ref{fig::4},a), generating at the defect core a net flow directed towards its head, as observed in MDCKs or progenitor cells \cite{saw2017topological,kawaguchi2017topological}. 
Remarkably, the vorticity map of $-1/2$ defects presents a three  fold organisation of pairs of counter-rotating vortices Fig.~(\ref{fig::4},b). These observations are in excellent agreement with the flow field of isolated defects in extensile active nematics \cite{giomi2014defect}. Besides, weak negative divergence is found near the cores of both type of defects Fig.~(\ref{fig::4},c,d), suggesting an accumulation of cells near these regions, which is not captured by incompressible flows (i.e. $\nabla\cdot\mathbf{v}=0$). 

In conclusion, we have shown that HBECs monolayers develop $2d$ turbulent collective flows featured by spontaneous emergence of mesoscopic vortices and nematic defects over a wide range of system's activity. The properties of cell flows and cell orientation are analysed for length scales of millimetres and time scales of tens of hours, granting access to high statistics with which the predicted behaviour of active nematics is tested. Our findings show that regardlessly of the instantaneous cell density, the statistics of vortices are compatible with the chaotic dynamics of active nematics. In particular we show that the total distribution of vortex area decays exponentially with a mean value that remains constant in time. Other biological systems has been found to exhibit similar features \cite{guillamat2017taming}. Furthermore, we analyse the spatiotemporal dynamics of cell orientation and their influence on local cell flows. We observe that our system exhibits spontaneous formation and annihilation of pairs of nematic defects with opposite charges, such that their instantaneous number remains constant. The vorticity of cell flows near $\pm1/2$ defects is in excellent agreement with flow fields generated by isolated defects in active liquid crystals. Comparing our findings with  active nematics, we deduce that the anisotropic active stresses and the Frank constant are weakly density-dependent in this system, whereas the effective shear viscosity may vary more than a decade. We believe that these cellular systems can serve as a testing ground to further explore the physical principles governing their mechanics.

\begin{acknowledgments}
We thank Jacques Prost and Jean-Fran\c cois Joanny for enlightening discussions and Julia Yeomans for suggesting the cell orientation analysis. We gratefully acknowledge the CelTisPhysBio Labex for financial support. L.G. is supported by the Netherlands Organization for Scientific Research (NWO/OCW) via the Frontiers of Nanoscience program and the Vidi scheme. The Biology-inspired physics at meso-scales group is member of the CelTisPhysBio Labex and of the Institut Pierre-Gilles de Gennes. 
\newline
C. B.-M. and V. Y. contributed equally to this work.
\end{acknowledgments}

\bibliography{apssamp.bib}

\end{document}